\documentstyle[aps,epsfig,prl]{revtex}
\begin{document}
\title{Reflection and absorption
  of QWs irradiated by light pulses in a strong magnetic field}
\author{I. G. Lang, L. I. Korovin}
\address{A. F. Ioffe Physical-Technical Institute, Russian
Academy of Sciences, 194021 St. Petersburg, Russia}
\author{D.A. Contreras-Solorio, S. T. Pavlov\cite{byline1}}
\address {Escuela de Fisica de la UAZ, Apartado Postal C-580,
98060 Zacatecas, Zac., Mexico}
\twocolumn[\hsize\textwidth\columnwidth\hsize\csname
@twocolumnfalse\endcsname
\date{\today}
\maketitle \widetext
\begin{abstract}
\begin{center}
\parbox{6in}
{It has been shown that the non-sinusoidal character oscillations
appear in the transmitted, reflected and absorbed light fluxes
when light pulses irradiate a semiconductor quantum well (QW),
containing a set of a large number of the equidistant energy
levels of electronic excitations. The oscillation amplitude is
comparable to the flux values  for the short pulses, duration of
which $\gamma_l^{-1} \leq \hbar/\Delta E$. A damping echo of the
exciting pulse appears through the time intervals
$2\pi\hbar/\Delta E$ in the case of the very short light pulses
$\gamma_l^{-1}\ll \hbar/\Delta E$. Symmetrical and asymmetrical
pulses with a sharp front have been considered. Our theory is
applicable for the narrow QWs in a strong magnetic field, when the
equidistant energy levels correspond to the electron-hole pairs
(EHP) with different Landau quantum numbers.}
\end{center}
\end{abstract}

]
 \narrowtext
\section{Introduction}

A response of different physical systems on a pulse light
irradiation is a subject of strong interest in the past decade
\cite{1,2,3,4}. The availability of short-pulse techniques and
commercial devices made it possible to investigate coherent
phenomena in the processes of interaction between the light and
elementary excitations in various systems, thus providing valuable
information on their excitation spectra and the mechanisms of
relaxation.

    A row of theoretical and experimental investigations is
devoted to the elaborate study of the Wannier-Mott excitons in the
bulk crystals and semiconductor QWs with the help of the time
resolved scattering (TRS), because just the existence of the
discrete energy levels determines the most interesting results
obtainable by the TRS. It is well known that a couple of the
closely disposed energy levels demonstrate a new effect: The
sinusoidal beatings appear in reflected and transited pulses on a
frequency corresponding to the energy distance between the energy
levels (see, for instance, Ref.~\onlinecite{1}).

        In this paper we examine theoretically light pulse
reflection and absorption by the semiconductor QWs in a strong
magnetic field (SMF) directed perpendicularly to the QW plane
$xy$. In such a case, excitations in the QW are characterized by
the quasi-momentum ${\bf K}_\perp$ in the $xy$ plane, because the
system is homogeneous in this plane. If a movement along the $z$
axis is a finite one, the rest of the indexes of the excitations
are discrete ones. When the excitations are created by light, the
condition ${\bf K}_\perp={\bf \kappa}_\perp$ , where
${\bf\kappa}_\perp$ is the light wave vector projection on the
$xy$ plane, is satisfied. We examine the case of the normal
irradiation when ${\bf K}_\perp={\bf\kappa}_\perp=0$. Under this
condition the spectrum is always discrete for the finite movement
along the $z$ axis.\cite{5}

        If the light pulse carrier frequency  $\omega_l$
exceeds slightly the QW semiconductor energy gap $E_g$,  then the
created by light excitations are the electron-hole pairs (EHP),
which are characterized  by the indexes $n_e=n_h=n,~ l_e,~ l_h$,
where $n_e (n_h)$ is the Landau quantum number of an electron
(hole), $l_e (l_h)$ is the electron (hole) size-quantization
(along the $z$ axis) quantum number. In the case of the infinitely
deep QW $l_e=l_h$, but we will not restrain our consideration by
this restriction. The excitation energy, measured from the ground
state energy, is equal  
\begin{equation}
\label{1}
E_{\xi_0}=E_g+\varepsilon^e_{l_e}+\varepsilon^h_{l_h}
+(n+1/2)\hbar\Omega_\mu,
\end{equation}
where $\xi_0$ is the set of the indexes ${\bf K}_\perp=0,~ n,~
l_e,~ l_h$;~ $\varepsilon^e_{l_e}~ (\varepsilon^h_{l_h})$ is the
electron (hole) energy on the size-quantized energy level $l$
(see, for instance, Ref.~\onlinecite{6}),~ $\Omega_\mu=|e|H/\mu
c,~ \mu=m_em_h/(m_e+m_h),~ m_e~ (m_h)$ is the electron (hole)
effective mass. It follows from Eq. (1) that the excitation
energies are equidistant for the fixed quantum numbers $l_e$ and
$l_h$. Generally speaking, the level equidistancy is broken if one
takes into account the Coulomb electron-hole interaction, i. e.
the excitonic effect. But  the Coulomb interaction is a weak
perturbation, if the conditions
$$d<<a,~~~ a_H<<a,$$
( $d$ is the QW width, $a$ is the Wannier-Mott exciton radius in
the magnetic field absence, $a_H=(c\hbar/|e|H)^{1/2}$ is the
magnetic length) are satisfied, and leads only to the small shifts
of the  energy levels Eq. (1).\cite{7} We suppose that the QWs are
narrow  and that magnetic fields are strong enough to ignore the
excitonic effects. An equidistancy violation may be due to the
 semiconductor band non-parabolicity , but we suppose that the
 non-parabolicity
is weak in the vicinity of the bottoms of the conduction and
valence bands.

    Let us suppose that a QW with a system of the equidistant
energy levels is irradiated by the light pulse with a carrier
frequency $\omega_l$. Let us admit that the carrier frequency
$\omega_l$ is alongside the resonance with one of the energy
levels. Then two variants are possible: One can ignore the
influence of the rest levels or take into account some number of
the neighbor energy levels. The choice between the variants
depends on the pulse form and duration, i. e. from the pulse
frequency spectrum.

Some ladder-type structure has been predicted in Refs.~\cite{8,9}
for the reflected and transmitted pulses for the QW in the SMF
irradiated by the sharply asymmetrical light pulse with a sharp
front, which corresponds to the second variant mentioned above.
The first variant can be realized, for instance, in the case of a
symmetrical pulse under condition $\gamma_l<<\Delta\omega$,~
$\Delta\omega$ is the distance between the neighbor levels,
$\gamma_l^{-1}$ is the pulse duration,  , $\gamma_l$ is the
frequency dispersion.

    A large number of the theoretical investigations have been
devoted to the examination of the QW electronic system response in
the cases of one or two excited energy levels. We have to stress
that a violation of the translation invariance  in the direction
perpendicular to the QW plane leads to the radiative broadening
$\gamma_r$ of the energy levels \cite{10,11}. In the case of the
high quality QWs the radiative broadening may be comparable to or
even exceed  the contributions of other relaxation mechanisms.
This physical situation demands an adequate theoretical
description, where one has to take into account the high orders of
the electron-electromagnetic field (EMF) interaction (see
Refs.~\cite{8,9,10,11,12,13,14,15,16,17,18,19,20,21}. In these
investigations a transparent QW is considered, i. e. it is
supposed, that light absorption and reflection are due to the
existence of one or two excited energy level. In the case when
\begin{equation}
\label{2}
\gamma_r<<\gamma,
\end{equation}
where $\gamma$ is the non-radiative inverse lifetime of the
electronic excitations, the  perturbation theory is applicable and
the lowest order on the electron-EMF interaction is enough when
the response on the monochromatic or pulse irradiation is
investigated.\cite{byline2}

     Under condition Eq. (2) the induced EMFs on the left- and on the right hand side
of the QW are small in comparison with the exciting EMFs; in the
case of monochromatic irradiation the reflection (${\cal R}$) and
absorption (${\cal A}$) coefficients are small in comparison with
unity, and the transmitted pulse differs slightly from the
exciting pulse form. However, even in this situation the very
interesting experimental results are obtained: Delaying of a short
pulse on the times of order  $\gamma^{-1}$ had been seen in the
transmitted light, and the sinusoidal beatings on the frequencies
$\Delta E/\hbar$ (where $\Delta E$ is the distance between the
energy levels) had been seen (see, for instance,
Ref.~\onlinecite{1}).

        In the opposite case
\begin{equation}
\label{3}
\gamma_r\ge\gamma,
\end{equation}
the induced EMFs are comparable to the exciting EMFs, the
coefficient ${\cal R}$ can be close to unity, the coefficient
${\cal A}$ is close to 1/2. The results of the QW irradiation by
the monochromatic light under condition Eq. (3) in the case of the
only excited energy level had been obtained in Refs.~
\onlinecite{10,11,12,13,18} and in the case of the two excited
energy levels in Ref.~\onlinecite{20}, respectively. The form of
the reflected and transmitted pulses closely to the resonance of
the carrier frequency with the only energy level in a QW had been
found in Ref.~\cite{17}. The analytical solution for a
non-symmetrical pulse with a sharp front had been found and the
numerical calculation results had been obtained for the
symmetrical "one-overhyperbolic-cosine" pulse. The EHP radiative
lifetimes in QWs in SMFs had been calculated in Refs.~\cite{8,20}.

    In this paper we study a response of a multilevel excitation
system in a QW, subjected to the SMF and irradiated by the
symmetric light pulse. The results are compared with those for the
symmetrical pulse with a sharp front.

\section{Electric fields on the right- , left
hand side from a QW irradiated by light pulses.}

    Let us assume that a time-limited light pulse drops down
from the left on a single QW perpendicular to its surface. The
electric field of the pulse is
\begin{eqnarray}
\label{4}
 {\bf E}_0(z,t)&=&E_0{\bf e}_1
e^{-i\omega_lp}\left\{\Theta(p)
e^{-\gamma_{l1}p/2}\right.\nonumber\\
&+&\left.[1-\Theta(p)] e^{\gamma_{l2}p/2}\right\}+c.c.,
\end{eqnarray}
where $E_0$ is the real amplitude, ${\bf e}_l$ is the polarization
vector, $\omega_l$ is the pulse carrier frequency ,
\begin{equation}
\label{5}
p=t-zn/c,
\end{equation}
$n$ is the refraction index out of the QW , $\Theta(p)$ is the
Haeviside function . The Umov-Pointing vector corresponds to the
pulse Eq.(4)
\begin{equation}
\label{6}
{\bf S}(z, t)={\bf S}_0P(p),
\end{equation}
\begin{eqnarray}
\label{7} {\bf S}_0={{\bf e}_z\over 2\pi}{c\over
n}E_0^2,\nonumber\\ P(p)=\Theta(p) e^{-\gamma_{l1}p/2}
+[1-\Theta(p)] e^{\gamma_{l2}p/2}\},
\end{eqnarray}
${\bf e}_z$ is the unit vector along the $z$ axis. Let us make the
Fourier transformation of Eq.(4)
\begin{equation}
\label{8}
 {\bf E}_0(z,t)=
E_0{\bf e}_1\int_{-\infty}^\infty d\omega
e^{-i\omega p}D_0(\omega)
+c.c.,
\end{equation}
\begin{equation}
\label{9}
D_0(\omega)={i\over 2\pi}\left[(\omega-\omega_l+i\gamma_{l1})^{-1}
-(\omega-\omega_l-i\gamma_{l2})^{-1}\right]
\end{equation}
The pulse is symmetrical one under condition
\begin{equation}
\label{10} \gamma_{l1}=\gamma_{l2}=\gamma_{l}.
\end{equation}
When $\gamma_l\rightarrow 0$ it transfers into a monochromatic
light wave with the frequency $\omega_l$, and the function 
$D_0(\omega)$ transfers into the Dirac  
$\delta(\omega-\omega_l)$-function.

        In Refs.~\cite{8,9,17} a strongly asymmetrical pulse with a sharp front
had been used, when $\gamma_{l2}\rightarrow\infty$ and the second
term in the braces in the RHS of Eq. (4) vanishes, as well as the
second term in the
 square brackets in the
RHS of Eq. (9).

        The pulse Eq. (4) is very useful for calculations. Its
imperfection is a sharp form of the peak at $t-zn/c=0$, i. e. the
derivative discontinuity of the function  $P(p)$ at $p=0$.
However, all the qualitative conclusions of the theory, obtained
below, do not change for the smooth pulses.

Let us suppose that the incident waves have the circular
polarization, i. e.
\begin{equation}
\label{11}
{\bf e}_l={1\over\sqrt{2}}({\bf e}_x\pm i{\bf e}_y),
\end{equation}
where ${\bf e}_x, {\bf e}_y$ are the unit vectors along the axis
$x, y$.

       Let us consider the QW , the width $d$ of which is much
smaller than the light wave length $c/n\omega_l$. Then the
electric field ${\bf E}_{left (right)}$ on the left (right) hand
side of the QW is determined by the expressions\cite{17}
\begin{equation}
\label{12}
{\bf E}_{left (right)}(z, t)={\bf E}_0(z, t)+\Delta
{\bf E}_{left (right)}(z, t),
\end{equation}
\begin{eqnarray}
\label{13} \Delta{\bf E}_{left (right)}(z, t)= E_0{\bf
e}_1\int_{-\infty}^\infty d\omega e^{-i\omega(t\pm
zn/c)}D(\omega)\nonumber\\ +c.c.,
\end{eqnarray}
where the upper (lower)sign corresponds to the index "left"
("right").  The frequency partition function  $D(\omega)$ is
determined as
\begin{equation}
\label{14}
{\cal D}(\omega)=-{4\pi\chi(\omega){\cal D}_0(\omega)\over
1+4\pi\chi(\omega)},
\end{equation}
\begin{eqnarray}
\label{15} \chi(\omega)&=&{i\over 4\pi}
\sum_\rho{\gamma_{r\rho}\over 2}\nonumber\\&\times&
[(\omega-\omega_\rho+i\gamma_\rho/2)^{-1}+
(\omega+\omega_\rho+i\gamma_\rho/2)^{-1}]\nonumber\\
&+&{Q(\omega)\over4\pi}-i{I(\omega)\over4\pi},
\end{eqnarray}
where the index $\rho$ is the number of the excited state,
$\hbar\omega_\rho$ is the excitation energy measured from the
ground state energy, $\gamma_{r\rho}(\gamma_\rho)$
 is the radiative (non-radiative) inverse lifetime of the excitation
$\rho$,~ $Q(\omega)$ and  $I(\omega)$ determine the contributions
to the real and imaginary parts of the function
 $\chi(\omega)$ due to the unaccounted electronic excitations (for instance,
the excitations from the deeper levels than the valence
 band) and by the lattice excitations.

 \underline{We will consider, as in Refs.~\cite{8,9,10,11,12,13,14,15,16,17,18,20},
 a transparent QW,} i. e. we presume
\begin{equation}
\label{16} Q(\omega)\simeq 0,~~~ I(\omega)\simeq 0,
\end{equation}
and ignore the second non-resonant term in the square brackets in
the  RHS of Eq.(15).\cite{byline3}
 Thus, we suppose
\begin{equation}
\label{17}
\chi(\omega)\simeq
{i\over 4\pi}
\sum_\rho{\gamma_{r\rho}\over 2}
(\omega-\omega_\rho+i\gamma_\rho/2)^{-1}.
\end{equation}
Substituting Eq. (17) into Eq. (14), we obtain
\begin{equation}
\label{18}
D(\omega)=-{i\sum_\rho(\gamma_{r\rho}/2)
(\omega-\omega_\rho+i\gamma_\rho/2)^{-1}D_0(\omega)\over
1+i\sum_\rho(\gamma_{r\rho}/2)
(\omega-\omega_\rho+i\gamma_\rho/2)^{-1}}.
\end{equation}
Substituting Eq. (18) into Eq. (13), we obtain the induced field
as a sum of two terms
\begin{equation}
\label{19} \Delta{\bf E}_{right}(z, t)=\Delta{\bf E}_1(p)+\Delta
{\bf E}_2(p).
\end{equation}
The first is the contribution of the poles of the function
$D_0(\omega)$. Applying
 Eqs. (9) and (13), it is easily to obtain
\begin{eqnarray}
\label{20} \Delta{\bf E}_1(p)&=&-iE_0{\bf
e}_le^{-i\omega_lp}\nonumber\\ &\times&
\left\{
\Theta(p)e^{-\gamma_{l1}p/2}\right.\nonumber\\ &\times&
{\sum_\rho(\gamma_{r\rho}/2)
[\omega_l-\omega_\rho+i(\gamma_\rho-\gamma_{l1})/2]^{-1}\over
1+\sum_\rho(\gamma_{r\rho}/2)
[\omega_l-\omega_\rho+i(\gamma_\rho-\gamma_{l1})/2]^{-1}}\nonumber\\
&+&[1-\Theta(p)]e^{\gamma_{l2}p/2}\nonumber\\
&\times&\left.{\sum_\rho(\gamma_{r\rho}/2)
[\omega_l-\omega_\rho+i(\gamma_\rho+\gamma_{l2})/2]^{-1}\over
1+\sum_\rho(\gamma_{r\rho}/2)
[\omega_l-\omega_\rho+i(\gamma_\rho+\gamma_{l2})/2]^{-1}}\right\}\nonumber\\
&+&c.c..
\end{eqnarray}
The second term $\Delta {\bf E}_2(p)$ is the contribution of the
poles of the function $$ {i\sum_\rho(\gamma_{r\rho}/2)
(\omega-\omega_\rho+i\gamma_\rho/2)^{-1}\over
1+i\sum_\rho(\gamma_{r\rho}/2)
(\omega-\omega_\rho+i\gamma_\rho/2)^{-1}}. $$
In the case of one excited energy level we have one pole
$\omega_0-i(\gamma+\gamma_r)/2$, in the case of two energy levels
we have two poles, the positions of which are easily determined.

However, already in the case of three energy levels it is
difficult to determine the poles, because one has to solve the
third order equation. Therefore to calculate  $\Delta{\bf E}_2(p)$
in the case of the large number of the energy levels  one needs to
use an approximation, which is applicable only for the small
values
 $\gamma_{r\rho}$.
Constraining with the lowest  order term on the electron-EMF
interaction and supposing

\begin{equation}
\label{21}
D(\omega)\simeq-4\pi\chi(\omega)D_0(\omega),
\end{equation}
we obtain the approximate result for the function  $\Delta{\bf
E}_2(p)$
\begin{eqnarray}
\label{22} \Delta{\bf E}_2(p)&\simeq&-iE_0{\bf e}_l \Theta(p)
\sum_\rho(\gamma_{r\rho}/2)e^{-i\omega_\rho
p-(\gamma_\rho)/2}\nonumber\\
&\times&\left\{[\omega_l-\omega_\rho+i(\gamma_\rho-\gamma_{l1})/2
]^{-1}\right.\nonumber\\
&-&\left.[\omega_l-\omega_\rho+i(\gamma_\rho+\gamma_{l2})/2
]^{-1}\right\}+c.c..
\end{eqnarray}

Eq. (20) contains the sum
 $$S=\sum_\rho(\gamma_{r\rho}/2)
(\omega_l-\omega_\rho+i\bar{\gamma}_\rho/2)^{-1},$$ where
$\bar{\gamma}_\rho=\gamma_\rho-\gamma_{l1}$~ or~
$\bar{\gamma}_\rho=\gamma_\rho+\gamma_{l2}$. The real parts of
these sums diverge, if the values
 $\gamma_{r\rho}$ do not depend on $\rho$.
In the case of the equidistant energy levels this divergence is
logarithmic one. This divergence is eliminated indeed, but it is
very difficult to determine the eliminating mechanism. Thus we
will act as follows: Let us write the sum $S$ as
\begin{equation}
\label{23} S=\sum_{\rho}\prime(\gamma_{\rho}/2)
(\omega_l-\omega_\rho+i\bar{\gamma}_\rho/2)^{-1}-J(\omega_l),
\end{equation}
where the sign $^\prime$ in the sum on $\rho$ means the summing on
the limited number of the energy levels. The non-dimensional
value  $J(\omega_l)$ depends weakly on $\omega_l$, if the
frequency set, determined by Eq.  (9), is in the resonance with
the level group, containing in the first term in the
 RHS of Eq. (23). Eq. (23)
has to be opposed with Eq. (15). The value  $J(\omega_l)\simeq
 J$ is added to the constant $I(\omega)\simeq I$.
 Thus, we do not know the value  $J$ , and in any case,~ $J<<1$~ or~
 $J\ge1$, the contribution $\Delta{\bf E}_1(p)$ to induced electric
 field damps on times of order $\gamma_l^{-1}$, as the follows from Eq. (20).
We postpone  the question about the divergencies ' which can
appear in the RHS of Eq.(22) until  section Y. We will see that
the divergencies do not appear in the case of the symmetrical
pulses. The expression for the induced field $\Delta{\bf
E}_{left}(z, t)$ on the left hand side of the QW differs from Eqs.
(19)-(22) only by substitution $s=t+zn/c$ instead of~
 $p=t-zn/c$.
\section{Transmitted, reflected and
absorbed energy fluxes.}

       For the sake of brevity
we will call the Umov-Pointing vectors as the energy fluxes. The
transmitted energy flux, i.e. the flux on the right hand side of
the QW, is equal
\begin{equation}
\label{24} {\bf S}_{right}(z, t)= {{\bf e}_z\over 4\pi}{c\over
n}|{\bf E}_{right}(z, t)|^2,
\end{equation}
the energy flux on the left hand side of the QW, is equal
\begin{equation}
\label{25} {\bf S}_{left}(z, t)={\bf S}(z, t)+{\bf S}_{ref}(z, t),
\end{equation}
where ${\bf S}(z, t)$ is the energy flux of the exciting pulse
determined by Eq. (6), ${\bf S}_{ref}(z, t)$ is the reflected
energy flux, which is equal
\begin{equation}
\label{26} {\bf S}_{ref}(z, t)=-{{\bf e}_z\over 4\pi}{c\over n}
|\Delta{\bf E}_{left}(z, t)|^2.
\end{equation}
The absorbed energy flux is defined as
\begin{equation}
\label{27} {\bf S}_{abs}(t)={\bf S}_{left}(z=0, t)-{\bf
S}_{right}(z=0, t)
\end{equation}
and equals
\begin{equation}
\label{28} {\bf S}_{abs}=-{{\bf e}_z\over 2\pi}{c\over n}{\bf
E}_{right} (z=0, t)\Delta{\bf E}(z=0, t),
\end{equation}
where
\begin{eqnarray}
\label{29} \Delta{\bf E}(z=0, t)&=&\Delta{\bf E}_{left}(z=0,
t)\nonumber\\
 &=&\Delta{\bf E}_{right}(z=0, t).
\end{eqnarray}
Let us introduce the non-dimensional functions
 ${\cal R}(t), {\cal A}$(t) and ${\cal T}(t)$ with the help of the
 interrelations
\begin{eqnarray}
\label{30} {\bf S}_{ref}(z, t)=-{\bf S}_0{\cal R}(s),~~~ {\bf
S}_{abs}(t)={\bf S}_0{\cal A}(t),\nonumber\\ {\bf S}_{right}(z,
t)={\bf S}_0{\cal T}(p).
\end{eqnarray}
It follows from Eq. (27), that
\begin{equation}
\label{31} {\cal R}(t)+{\cal A}(t)+{\cal T}(t)=P(t).
\end{equation}
The values ${\cal R}(t)$ and ${\cal T}(t)$ are always positive
ones, the absorption  ${\cal A}(t)$ may be positive as well as
negative.

       Let us consider a system with the arbitrary number of the
energy levels irradiated with the pulse Eq. (4). When calculating
Eq. (22) we have already used the condition, according to which
the parameters $\gamma_{r\rho}$ are the smallest ones among the
parameters of our task, and the times
 $p<<\gamma_{r, \rho}^{-1},~ s<<\gamma_{r, \rho}^{-1}$
have been considered.  Now we use an additional condition of a
short pulse
\begin{equation}
\label{32} \gamma_\rho<<\gamma_l
\end{equation}
and consider the times 
\begin{equation}
\label{33} p>>\gamma_l^{-1},~~~ s>>\gamma_l^{-1}.
\end{equation}
It is obvious, that then only the contribution
 $\Delta{\bf E}_2(p)$, containing,according to Eq. (22),
$\exp (-\gamma_\rho p/2)$, is preserved in the  RHS of Eq. (19).
The contribution $\Delta{\bf E}_1(p)$, determined in Eq. (20), ¨
and the exciting field ${\bf E}_0(z, t)$ are small, because they
content the factor
 $\exp (-\gamma_lp/2)$.
 Thus, under conditions Eqs. (32) and (33) we obtain
\begin{equation}
\label{34} {\bf E}_{right}(z, t)\simeq\Delta{\bf E}_2(p),~~~ {\bf
E}_{left}(z, t)\simeq\Delta{\bf E}_2(s),
\end{equation}
from which it follows, that the transmitted and reflected fluxes
are equal in absolute values, i. e.
\begin{equation}
\label{35} {\cal R}(t)\simeq{\cal T}(t).
\end{equation}
Because on the times of Eq. (33) $$P(t)\simeq 0,$$ we obtain from
Eq. (31)
\begin{equation}
\label{36} {\cal A}(t)=-2{\cal R}(t).
\end{equation}
The negative absorption, which is equal to the double reflection,
means that the QW gives back the accumulated energy, radiating it
symmetrically by two fluxes on the left- and on the right hand
side of the QW. This energy was accumulated as excitations,
created during the pulse transmission.
\section{The equidistant energy
level system in a QW in a SMF.}

Under conditions Eqs. (32) and (33) in the case of the symmetrical
pulse we obtain from Eqs. (34) and (22)
\begin{eqnarray}
\label{37} {\bf E}_{right}^{sim}(z, t)&\simeq&iE_0{\bf e}_l
\Theta(s) \sum_\rho(\gamma_{r\rho}/2) e^{-i\omega_\rho
p-\gamma_\rho p/2}\nonumber\\
&\times&\left\{[\omega_l-\omega_\rho+i(\gamma_\rho-\gamma_{l})/2]^{-1}
\right.\nonumber\\
&-&\left.[\omega_l-\omega_\rho+i(\gamma_\rho+\gamma_{l})/2]^{-1}\right\}.
\end{eqnarray}
In the case of the asymmetrical pulse with the sharp front (see
Refs.~ \cite{8,9,17} we obtain the result of Eq. (37), in which
the second term in the braces is absent. One can see easily, that
the sum on $\rho$ in the RHS of Eq. (37) does not diverge at the
large values of  $\rho$, no matter what  the value  of the
variable $p$ is. This divergence can appear in the case of an
asymmetrical pulse.

 Let us simplify Eq. (37), supposing that the inverse lifetimes
$\gamma_{r\rho}$ and $\gamma_\rho$ are equal for all the energy
levels, i. e.
\begin{equation}
\label{38} \gamma_{r\rho}\simeq\gamma_r,~~~
\gamma_\rho\simeq\gamma,
\end{equation}
and ignoring the small values $\gamma$ comparing with $\gamma_l$
in the square brackets. Then for the reflected flux and the
symmetrical pulse we obtain with the help Eq. (26)
\begin{eqnarray}
\label{39} {\cal R}^{sim}(s)&\simeq&{\gamma_r^2\gamma_l^2\over
4}e^{-\gamma s} \sum_{\rho, \rho^\prime}
[(\omega_l-\omega_\rho)^2+\gamma_l^2/4]^{-1}\nonumber\\
&\times&[(\omega_l-\omega_{\rho^\prime})^2+\gamma_l^2/4]^{-1} \cos
[(\omega_\rho-\omega_{\rho^\prime})s].
\end{eqnarray}
The analogical result for the asymmetrical pulse is as follows
\begin{eqnarray}
\label{40} {\cal R}^{asim}(s)&\simeq&{\gamma_r^2\over 4}e^{-\gamma
s} \sum_{\rho, \rho^\prime}
[(\omega_l-\omega_\rho)^2+\gamma_l^2/4]^{-1}\nonumber\\
&\times&[(\omega_l-\omega_{\rho^\prime})^2+\gamma_l^2/4]^{-1}\nonumber\\
&\times&\{(\gamma_l/2)(\omega_\rho-\omega_{\rho^\prime}) \sin
[(\omega_\rho-\omega_{\rho^\prime})s]\nonumber\\
&+&[(\omega_l-\omega_\rho)(\omega_l-\omega_{\rho^\prime)}+\gamma_l^2/4]
\nonumber\\ &\times&\cos [(\omega_\rho-\omega_{\rho^\prime})s]\}.
\end{eqnarray}

Let us apply Eqs. (39), (40) in the case of the equidistant Landau
levels (LLs), corresponding to Eq. (1) for the EHP energy at the
fixed size-quantized numbers $l_e$ and $l_h$ for electrons and
holes, respectively. The frequencies $\omega_\rho$ and $\omega_l$
from the RHS of Eqs. (39),(40) are measured from the value
$$E_g/\hbar+\varepsilon^e_{le}+\varepsilon^h_{lh}+\Omega_\mu/2.$$
Then
\begin{equation}
\label{41} \omega_\rho=n\Omega_\mu,~~~    n=0, 1, 2........
\end{equation}

Let us substantiate the suppose Eq. (38) for the QW in a SMF. The
radiative lifetime of the EHPs, which is applicable for the heavy
holes in GaAs, has been calculated in Ref. \onlinecite{20}. There
are two sorts of the EHPs, designated by the indexes I and II.
These sorts are distinguished by the values of the interband
momentum matrix elements ${\bf p}_{cv}$, which are
\begin{equation}
\label{42} {\bf p}_{cv}^{I}={1\over\sqrt{2}}p_{cv}({\bf e}_x-i{\bf
e}_y),~~~ {\bf p}_{cv}^{II}={1\over\sqrt{2}}p_{cv}({\bf e}_x+i{\bf
e}_y).
\end{equation}
When we use the circular polarizations Eq. (11), every
polarization is linked strongly with the EHP sort I or II, because
the EHP-light interaction is proportional to the scalar production
${\bf e}_l{\bf p}_{cv}$ (compare with \cite{byline3}).
 It has been shown\cite{20}that the inverse radiative
lifetime of any sort is equal at ${\bf K}_\perp=0$
\begin{equation}
\label{43} \gamma_{r\xi_0}=2{e^2\over c\hbar}{\Omega_0\over n}
{p_{cv}^2\over m_0E_{\xi_0}} \pi^2_{l_e, l_h},
\end{equation}
where $\Omega_0=|e|H/m_0c$ is the cyclotron frequency for the bare
electron mass $m_0$, $n$ is the index of refraction, the energy
$E_{\xi_0}$ is determined in Eq. (1),
\begin{equation}
\label{44} \pi_{l_e, l_h}=\int_{-\infty}^{\infty}dz
\varphi_{cl_e}(z)\varphi_{vl_h}(z),
\end{equation}
$\varphi_{cl}(z),~ \varphi_{vl}(z)$ are the real functions,
corresponding to the size-quantization number $l$ for the QW of a
finite depth (see, for instance, Ref.~{6}), $c(v)$ is the
conduction (valence) band index.

 It
follows from Eq. (43) that the inverse lifetime $\gamma_{r\xi_0}$
is proportional to $H$. The dependence on the index  $n$ and the
magnetic field strength $H$ due to the factor $E_{\xi 0}$ in the
denominator is very weak, because in the RHS of Eq. (1) the energy
gap is much more than the energy $(n+1/2)\hbar\Omega_\mu$. We
suppose that there are no reasons for the strong dependence of
$\gamma_\rho$ on an integer $n$.

      With the help of Eq. (39) we obtain for the reflected flux
\begin{equation}
\label{45} {\cal
R}^{sim}(s)=4\left({\gamma_r\over\gamma_l}\right)^2e^{-\gamma s}
Y^{sim}_{\Omega_l, G_l}(S),
\end{equation}
where
\begin{eqnarray}
\label{46} Y^{sim}_{\Omega_l, G_l}(S)&=&{G_l^4\over 16}
\left\{\left[\sum_{n=0}^{\infty}{\cos(nS)\over
(\Omega_l-n)^2+G^2_l/4} \right]^2\right.\nonumber\\ &+&
\left.\left[\sum_{n=1}^{\infty}{\sin(nS)\over
(\Omega_l-n)^2+G^2_l/4} \right]^2\right\}.
\end{eqnarray}
The non-dimensional variables $S=\Omega_\mu s$ and
\begin{equation}
\label{47} \Omega_l={\omega_l\over\Omega_\mu},~~~
G_l={\gamma_l\over\Omega_\mu}.
\end{equation}
are introduced. The function Eq. (46) is periodical one with the
period $2\pi$ and symmetrical relatively the substitution of $S$
by $2\pi-S$.

In the case of the asymmetrical pulse we obtain instead of Eq.
(46) the expression which has a convenient form

\begin{eqnarray}
\label{48} Y^{asim}_{\Omega_l, G_l}(S)&=& {G_l^2\over
16}[(\sigma_{c0}\Omega_l-\sigma_{c1})^2
+(\sigma_{s0}\Omega_l-\sigma_{s1})^2]\nonumber\\ &+& {G_l^3\over
16}(\sigma_{s1}\sigma_{c0}-\sigma_{s0}\sigma_{c1}) + {G_l^4\over
64}(\sigma_{c0}^2+\sigma_{s0}^2),
\end{eqnarray}
where
\begin{eqnarray}
\label{49} \sigma_{c0}=
\sum_{n=0}^\infty{\cos(nS)\over(\Omega_l-n)^2+G_l^2/4},\nonumber\\
\sigma_{c1}=
\sum_{n=0}^\infty{n\cos(nS)\over(\Omega_l-n)^2+G_l^2/4},\nonumber\\
\sigma_{s0}=
\sum_{n=0}^\infty{\sin(nS)\over(\Omega_l-n)^2+G_l^2/4},\nonumber\\
\sigma_{s1}=
\sum_{n=0}^\infty{n\sin(nS)\over(\Omega_l-n)^2+G_l^2/4}.
\end{eqnarray}
\section{A symmetrical
exciting pulse. An echo in transmitted and reflected fluxes.}
 In the resonance of the frequency $\omega_l$ with a Landau
level, i. e. at

\begin{equation}
\label{50} \Omega_l=n_0,~~~ n_0=0, 1, 2,\ldots
\end{equation}
we obtain from Eq. (46)
\begin{eqnarray}
\label{51} Y^{sim}_{\Omega_l=n_0, G_l}(S)&=&{G_l^4\over
16}\nonumber\\
&\times&
\left\{\left[\sum_{n=0}^{\infty}{\cos(nS)\over n^2+G^2_l/4}
+\sum_{n=1}^{n_0}{\cos(nS)\over n^2+G^2_l/4}
\right]^2\right.\nonumber\\
&+&\left.\left[\sum_{n=n_0+1}^{\infty}
{\sin(nS)\over n^2+G^2_l/4}\right]^2\right\}.
\end{eqnarray}
In the extreme case
\begin{equation}
\label{52} \Omega_l=n_0,~~~ G_l<<1
\end{equation}
we obtain from Eq. (51)
\begin{equation}
\label{53} Y^{sim}_{\Omega_l=n_0, G_l<<1}(S)\simeq 1+ {G_l^2\over
2} \left[F_{sim}(S)+\sum_{n=1}^{n_0}{\cos(nS)\over n^2}\right],
\end{equation}
where
\begin{equation}
\label{54} F_{sim}(S)=\sum_{n=1}^\infty{\cos(nS)\over
n^2}={(\pi-S)^2\over 4}- {\pi^2\over 16}
\end{equation}
changes from $0$ to $2\pi$. At $G_l\rightarrow 0$ we obtain
$$Y_{\Omega_l=n_0, G_l=0}=1\ldots $$ and, according to Eq. (45),
$${\cal R}^{sim}(s)\simeq 4(\gamma_r/\gamma_l)^2e^{-\gamma
s}\cdots ,$$  which corresponds to the contribution from only one
LL, which is in the resonance with the frequency   $\omega_l$.

Further let us consider the case when the frequency $\omega_l$ is
in the resonance with one of the upper LLs, i. e.
\begin{equation}
\label{55} \Omega_l=n_0,~~~ n_0>>1,
\end{equation}
 and $G_l$ is arbitrary. Then we obtain immediately from Eq.
 (51)
\begin{equation}
\label{56} Y^{sim}_{\Omega_l=n_0, n_0>>1, G_l}(S)\simeq
{G_l^4\over 16} \left({4\over G_l^2}+
2\sum_{n=1}^{\infty}{\cos(nS)\over n^2+G^2_l/4}\right)^2.
\end{equation}
The sum in the round brackets in the RHS of Eq. (56) is calculated
precisely and we obtain
\begin{eqnarray}
\label{57} Y^{sim}_{\Omega_l=n_0, n_0>>1, G_l}(S)&\simeq&
{G_l^2\over 4}\pi^2\cosh^2[(\pi-S){G_l\over 2}]\nonumber\\
&\times&cosech^2\left( {\pi G_l\over 2}\right).
\end{eqnarray}
If $G_l<<1$ we obtain from Eq. (57)
\begin{equation}
\label{58} Y^{sim}_{\Omega_l=n_0, n_0>>1, G_l<<1}(S)\simeq
1+G_l^2F_{sim}(S),
\end{equation}
which accords with Eq. (53) in the limit $n_0\rightarrow\infty$.
Under condition
\begin{equation}
\label{59} G_l>>1
\end{equation}
it follows from Eq. (57)
\begin{eqnarray}
\label{60} &Y&^{sim}_{\Omega_l=n_0, n_0>>1, G_l>>1}(S)\simeq
{G_l^2\over 4}\nonumber\\&\times&\pi^2
\left\{\begin{array}{ll}e^{-SG_l},~~~ S<<1, &\quad\\
e^{-(2\pi-S)G_l},~~~ 2\pi-S<<1.\end{array} \right.
\end{eqnarray}

       In Figs. 1-4 the curves corresponding to Eq. (46)
are represented for the different values of the parameters
$\Omega_l$ and $G_l$. The functions $Y^{sim}_{\Omega_l, G_l}(S)$
are periodical ones with the period $2\pi$, only one period is
represented. Fig. 1 corresponds to the small value $G_l=0.1$ and
to the values $\Omega_l= 0;~ 0.05;~ 0.1$. At $\Omega _l=0$ the
frequency  $\omega_l$ is in the resonance with the lower LL for
the EHP;
 the values $\Omega_l=0.05$ and $\Omega_l=0.1$
correspond to the small deviation from this resonance. The curves
{\it a} are calculated on the precise formula Eq. (46) ,the curve
{\it b} is calculated on the approximate formula Eq. (53). One can
conclude from Fig.1 that in the case of the precise resonance of
the frequency $\omega_l$ with some  LL at small values $G_l$ the
periodical oscillations of the reflected and transmitted fluxes
have very small amplitudes and that the approximate formula Eq.
(53) gives the result, which is close to being precise. A small
deviation of the frequency $\omega_l$ from the resonance results
in a sharp drop of the flux values.

       Fig. 2 shows the same as Fig. 1, but at the bigger value
 $G_l=0.5$. In comparison with Fig. 1 the amplitude of the
 periodical beatings increases strongly, the approximate formula
Eq. (53) works worse , the small deviations of the frequency
$\omega_l $ from the resonance do not lead to the sharp decrease
of the energy fluxes.

       Fig. 3 corresponds to the parameter  $G_l=1$
and to the set of values $\Omega_l= 0;~ 0.1;~ 0.5;~ 1.0;~ 1.5;~
2.0.$ In this figure the amplitude of the periodical oscillations
of the non-dimensional factor $Y^{sim}_{\Omega_l, G_l}(S)$ reaches
the values 1.5 - 2.0. At $\Omega_l=0.5$ and $\Omega_l=1.5$,
 i. e. when the frequency $\omega_l$
is between the LLs $n=0$ and $n=1$  and between the LLs $n=1$ and
$n=2$, the curves touch the abscise axis at the point $S=\pi$,
thus, the fluxes approach to 0.

    Finally, Fig.4 corresponds to the large value
$G_l=5$ and to the set of values : $\Omega_l= 0;~ 0.1;~ 0.5;~
1.0.$ The values of the factor  $Y^{sim}_{\Omega_l, G_l}(S)$ at
points  $S=0$ and $S=2\pi$ grow sharply in comparison with the
corresponding values in Figs. 1-3, reaching 30,
 but they become very small in the interval
$S>>G_l^{-1}, (2\pi-S)>>G_l^{-1}$. Thus, the periodical function
$Y^{sim}_{\Omega_l, G_l}(S)$ is a sequence of the short pulses,
displaced with the interval $2\pi$; the duration of every pulse is
of order  $G_l^{-1}$. Applying Eq. (45) we obtain that at
$\gamma_l>>\Omega_\mu$ in the reflected energy flux there appear
echoes of the exciting pulse with the interval $2\pi/\Omega_\mu$,
which damp as $e^{-\gamma s}$. According to Eq. (35), the echo
would be observable in the transmitted flux also. For the values
of $\Omega_l$ from 0 to several units , when the frequency
$\omega_l$ is in the vicinity of the lower LLs, the echo's form is
distorted in the pulse replicas. In the resonance of $\omega_l$
with the upper LLs $n_0>>1$ we obtain, according to Eqs. (35),
(60), that the form of the echo-pulses coincides with the exciting
pulse form. But the values of these pulses is much smaller due to
the small factor $\pi^2(\gamma_r/\Omega_\mu)^2e^{-\gamma s}$.
\section{An asymmetrical exciting pulse.}

        Considering Eq. (48),
we find that this function diverges at points $S=2\pi m$, because
the sum $\sigma_{c1}$, defined by Eq. (49), diverges. It means
that for an asymmetrical exciting pulse we would obtain the
infinite values $Y^{asim}(S)$ in the points $S=0$ and $S=2\pi$
 in the Figs. 1-4. Of course this conclusion is wrong, because the
 values ${\cal R,~ A,~ T}$, defined by Eq. (30), cannot exceed unity.
Indeed, the infinite values are cut, but it is difficult to
determine the mechanism of this cutting. The approximation Eq.
(21) ( with the help of which we calculate $\Delta{\bf E}_2(p)$
from the  RHS of Eq. (19) in the lowest order on the
electron-light interaction) is inapplicable in the vicinity
$S=2\pi m$.

Therefore we include here only results, corresponding to the
resonant case, when the frequency $\omega_l$ coincide with one of
the upper Landau levels and the divergences do not appear. At
$\Omega_l=n_0, n_o>>1$ we obtain from Eq. (48)
\begin{eqnarray}
\label{61} Y^{asim}_{\Omega_l=n_0, n_0>>1, G_l}(S)&=& {G_l^2\over
16} \left [\sum_{n=-\infty}^\infty {n\sin(nS)\over n^2+G_l^2/4}
\right.\nonumber\\
 &+&\left.{G_l\over 2}\sum_{n=-\infty}^\infty
{n\cos(nS)\over n^2+G_l^2/4}\right]^2.
\end{eqnarray}
The sums from the RHS of Eq. (61) are calculated precisely, and we
obtain
\begin{equation}
\label{62} Y^{asim}_{\Omega_l=n_0, n_0>>1, G_l}(S)= {G_l^2\over
16}\pi^2 \left [cosech({\pi G_l\over 2})\right]^2e^{(\pi-S)G_l}.
\end{equation}
In the case  $G_l<<1$  we obtain from Eq. (62)
\begin{eqnarray}
\label{63} Y^{asim}_{\Omega_l=n_0, n_0>>1, G_l<<1}(S)
&\simeq&{1\over 4}[1+2G_lF(S)],\nonumber\\
F(S)&=&\sum_{n=1}^\infty{\sin(nS)\over n}={\pi-S\over 2}.
\end{eqnarray}
The result at $G_l=0$ is as following
\begin{equation}
\label{64} Y^{asim}_{\Omega_l=n_0, G_l=0}(S) \simeq{1\over 4},~~~
{\cal R}^{asim}(s)=(\gamma_r/\gamma_l)^2e^{-\gamma s},
\end{equation}
it corresponds to the resonance of the frequency $\omega_l$ with
only level in a QW. At $G_l>>1$ we have
\begin{equation}
\label{65} Y^{asim}_{\Omega_l=n_0, n_0>>1, G_l>>1}(S)
\simeq{\pi^2\over 4}G_l^2e^{-SG_l}.
\end{equation}
The last result means that in the case of the asymmetrical pulse
 $\gamma_l>>\Omega_\mu$ ,i.e. for very short pulses, the exciting
 pulse echo appears also. Under condition $n_0>>1$ the form of the reiterative pulses
coincide with the form of the exciting pulse, but the value of the
echo-pulse contains the small factor
$\pi^2(\gamma_r/\Omega_\mu)^2e^{-\gamma s}$, as well as in the
case of the symmetrical pulse.

Fig. 5 shows the function $Y^{asim}_{\Omega_l=n_0, n_0>>1,
G_l>>1}(S)$   (see Eq. (62)) at the values
 $G_l=0.1;~ 1.0;~ 10.$ At $G_l=0.1$ we obtain a saw-like curve and
a duplication of the exciting pulse form at $G_l=10$.

\section{ Conclusion}

Thus, we have calculated the time dependence of the transmitted,
reflected and absorbed energy fluxes, appearing under the normal
irradiation a QW, subjected to the SMF,  by the exciting light
pulse. We supposed that the energy levels of the electronic
excitations are equidistant with the energy interval
$\hbar\Omega_\mu$. The results for the symmetrical and
asymmetrical exciting pulses have been obtained. The
interrelations between the parameters have been chosen :
\begin{equation}
\label{66} \gamma_r<<\gamma,~~~ \gamma<<\gamma_l,~~~
\gamma<<\Omega_\mu,
\end{equation}
the interrelation of $\gamma_l$ and$\Omega_\mu$ is arbitrary one.
The energy fluxes have been examined on times
\begin{equation}
\label{67} t>>\gamma_l^{-1},~~~ t<<\gamma_r^{-1},
\end{equation}
when the exciting pulse is already damped, the transmitted and
reflected fluxes are equal in absolute values, and the absorbed
flux is negative  and equal in its modulus to the doubled
transmitted (or reflected) flux.

       The reflected flux contains the factor $e^{-\gamma s}$,
which determines its damping. Besides, there is the factor $Y$,
periodical on~ $s$~ with the period $2\pi\Omega_\mu^{-1}$. These
oscillations are never sinusoidal ones, characteristically only
for the case of two closely displaced excitation energy levels.
Under condition $\gamma_l<<\Omega_\mu$ , i.e. for the
comparatively short pulses, which time duration  exceeds the value
$\Omega_\mu^{-1}$, the oscillation amplitude is small.  In the
limit $\gamma_l/\Omega_\mu=0$ in the case of the resonance of the
frequency   $\omega_l$ with one of the energy levels, we obtain
the results, corresponding to the case of the only energy level.
Under condition $\gamma_l\ge\Omega_\mu$ the oscillation amplitude
becomes large one. Finally, in the case when
$\gamma_l>>\Omega_\mu$, i. e. the pulse duration is much smaller
than $\Omega_{\mu}^{-1}$, the damping echo of the exciting pulse
appears with the time intervals  $2\pi\Omega_\mu^{-1}$.

\section{Acknowledgements}
        S.T.P thanks the Zacatecas University and the National
Council of Science and Technology (CONACyT) of Mexico for the
financial support and hospitality. D.A.C.S. thanks CONACyT
(27736-E) for the financial support.
       Authors are grateful to
 A. D'Amore for a critical reading of the
manuscript.
       This work has been partially supported by the Russian
Foundation for Basic Research and by the Program "Solid State
Nanostructures Physics".

\begin{figure}

\caption{The function $Y^{sim}_{\Omega_l, G_l}(S)$, corresponding
to the periodical factor in the value of the reflected energy flux
when a QW is irradiated by the symmetrical light pulse. The pulse
duration $\gamma_l^{-1}$ exceeds on the order the value
$\Omega_\mu^{-1}$. }

\end{figure}

\begin{figure}

\caption{Same as Fig. 1, the pulse duration $\gamma_l^{-1} =2
 \Omega_\mu^{-1}$.
}

\end{figure}

\begin{figure}

\caption{Same as Fig. 1, the pulse duration  $\gamma_l^{-1}$=
 $\Omega_\mu^{-1}$.
}

\end{figure}

\begin{figure}

\caption{ Same as Fig.1 for the very small value $\gamma_l^{-1}$ ,
when the echo of the exciting pulse appears . }

\end{figure}

\begin{figure}

\caption{The function $Y^{asim}_{\Omega_l=n_0, n_0>>1, G_l}(S)$ in
the case of the asymmetrical exciting pulse with the sharp front.
The frequency
 $\omega_l$ is in the resonance with one of the upper Landau
 levels.
The values  $G_l=\gamma_l/\Omega_\mu$ are indicated. }

\end{figure}

\newpage
\begin{thebibliography}{99}
\bibitem[\dag]{byline1}On leave from the P. N. Lebedev Physical
Institute, Russian Academy of Sciences, 117924 Moscow, Russia.
\bibitem[\dag\dag]{byline2}
In the case of the pulse irradiation the lowest approximation on
the electron-EMF interaction is acceptable under condition Eq. (2)
only on the times  $t<<\gamma_r^{-1}$, because at
$t\ge\gamma_r^{-1}$ the intensity of the transmitted and reflected
light damps as $\exp(-\gamma_rt)$.
\bibitem[\ddag]{byline3}
Eqs. (13)-(15) at $Q(\omega)\simeq0, ~I(\omega) \simeq 0$ are
applicable, if each of the circular polarizations corresponds to
one of two types of the EHPs, the energies of which are equal.
\bibitem{1} H. Stolz, Time - Resolved Light Scattering from Exitons, Springer Tracts in Modern
Physics, Springer, Berlin,1994.
\bibitem{2} J.Shah, Ultrafast Spectroscopy of
Semiconductors and Semiconductor Nanostructures,
Springer, Berlin, 1996.
\bibitem {3} H. Hang, S. W. Koch, Quantum Theory of the
Optical and Electronic Properties of Semiconductors,
World Scientific, 1993.
\bibitem{4} S. Mukamel, Principles of Nonlinear Optical
Spectroscopy, Oxford University Press,NY, Oxford, 1995.
\bibitem{5} L. I. Korovin, I. G. Lang. S. T. Pavlov,Zh. Eksp. i Teor. Phys.,
{\bf 118}, issue 2(8), (2000).
\bibitem{6} I. G. Lang,V. I. Belitsky,
A. Cantarero, L. I. Korovin,   S. T.
Pavlov, M. Cardona,Phys. Rev. B, {\bf 54}, 17768 (1996).
\bibitem{7}I. V. Lerner, Yu. E. Lozovik, Zh. Eksp. i Teor. Phys., {\bf 78}, 1167 (1980).
\bibitem{8} I. G. Lang, V. I. Belitsky, M. Cardona,
Phys. stat. sol. (a), {\bf 164} 307 (1997).
\bibitem{9} I. G. Lang, V. I. Belitsky.
Physics Letters A, 245 , 329 (1998).
\bibitem{10} L. C. Andreani, F. Tassone, F. Bassani.
Solid. State Commun. , 77, 641 (1991).
\bibitem{11} L. C. Andreani, in Confined Electrons and Photons,
edited by E. Burstein and C. Weisbuch,
Plenum Press, NY, 1995, p. 57.
\bibitem{12} E. L. Ivchenko,
Sov. Phys. Solid State, {\bf 33}, 1344 (1991).
\bibitem{13}  F. Tassone, F. Bassani, L. C. Andreani,
Phys. Rev. B, {\bf 45}, 6023 (1992).
\bibitem{14} T. Stroucken et all, Phys. Rev. Lett. , {\bf 74}, 2391 (1995).
\bibitem{15} T. Stroucken et all, Phys. Rev. B , {\bf 53}, 2026 (1996).
\bibitem{16} M. Hubner et. al.,
Solid. State Commun. , {\bf 105}, 105 (1998).
\bibitem{17} I. G. Lang, V. I. Belitsky.
Solid State Commun. 107, 577 - 582 (1998).
\bibitem{18} L. C. Andreani, G. Panzarini, A. V. Kavokin, M. R. Vladimirova.
Phys. Rev. B {\bf 57}, 4670 (1998).
\bibitem{19} I. G. Lang, L. I. Korovin, A. Contreras Solorio
 and S. T. Pavlov, LANL archiv/cond-mat/0001248 .
 \bibitem{20}D. A. Contreras Solorio, S. T. Pavlov, , L. I. Korovin
 and I. G. Lang, LANL archiv/cond-mat/0002229 .
 \bibitem{21} L. I. Korovin, I. G. Lang,  A. Contreras Solorio
 and S. T. Pavlov, Fiz. Tverd. Tela, {\bf 42}, issue 12, p. 119
 (2000);\\ LANL archiv/cond-mat/0006364.
\end{thebibliography}
\end{document}